\documentstyle[sprocl]{article}

\input{psfig}

\def\be{\begin{equation}}
\def\ee{\end{equation}}
\def\bea{\begin{eqnarray}}
\def\eea{\end{eqnarray}}

\begin{document}

\title{\vspace*{-1.5cm}
\hspace*{\fill}{\normalsize LA-UR-98-3788} \\[1.5ex]
BOSE-EINSTEIN CORRELATIONS AND THE EQUATION OF STATE
OF NUCLEAR MATTER IN RELATIVISTIC HEAVY-ION COLLISIONS}

\author{B. R. Schlei}

\address{Physics Division P-25, Los Alamos National Laboratory,
Los Alamos,\\NM 87545, USA\\E-mail: schlei@LANL.gov} 


\maketitle\abstracts{Experimental spectra of the CERN/SPS 
experiments NA44 and NA49 are fitted while using four different 
equations of state of nuclear matter within a relativistic 
hydrodynamic framework. For the freeze-out temperatures, $T_f = 139$ MeV 
and $T_f = 116$ MeV, respectively, the corresponding freeze-out 
hypersurfaces and Bose-Einstein correlation functions for 
identical pion pairs are discussed. It is concluded, that the
Bose-Einstein interferometry measures the relationship between the
temperature and the  energy density in the equation of state of 
nuclear matter at the late hadronic stage of the fireball 
expansion. It is necessary, to use the detailed detector 
acceptances in the calculations for the Bose-Einstein correlations.}

\section{Introduction}

The equation of state (EOS) of nuclear matter at very high energy
densities is heretofore unknown. Many observables have been
proposed as a signature for a possible new state of nuclear matter, i.e.,
the quark-gluon plasma (QGP), which could be formed within the very
hot and dense zones of nuclear matter -- the fireballs --
in relativistic heavy-ion collision experiments.
Among the proposed signatures for a QGP are Bose-Einstein correlations 
(BEC), which are correlations of identical hadron pairs. BEC functions
are sensitive to the space-time dynamics of the fireball and therefore 
should give clues about the EOS, which governs the evolution of those
fireballs.

Among the many scientific contributions on BEC (for a recent overview, 
{\it cf.}, the book by Weiner \cite{wei97}) only few of them ({\it cf}., 
e.g., Rischke {\it et al.} \cite{ris96}) address the particular role 
of the EOS of nuclear matter in the intensity interferometry of 
identical hadrons, i.e., BEC.
It is the purpose of this paper to further investigate the
interplay between the EOS and BEC. In the following, we shall
consider a framework of analysis which is based on relativistic
hydrodynamics, because this approach allows for an explicit use of
an EOS. 

In order to address the issues raised above, we shall use
the simulation code HYLANDER-C \cite{brs97_2} with different equations 
of state and Cooper-Frye \cite{coo75} freeze-out. 
Experimental single particle momentum
distributions from 158$A$ GeV Pb+Pb collisions, measured by the NA44 
\cite{nxu96,bea96} and NA49  \cite{jon96} Collaborations, are used
as fit criteria for the hydrodynamical (one-fluid-type) calculations.

\begin{figure}
\vspace*{-2.0cm}
\begin{center}\mbox{ }
\psfig{figure=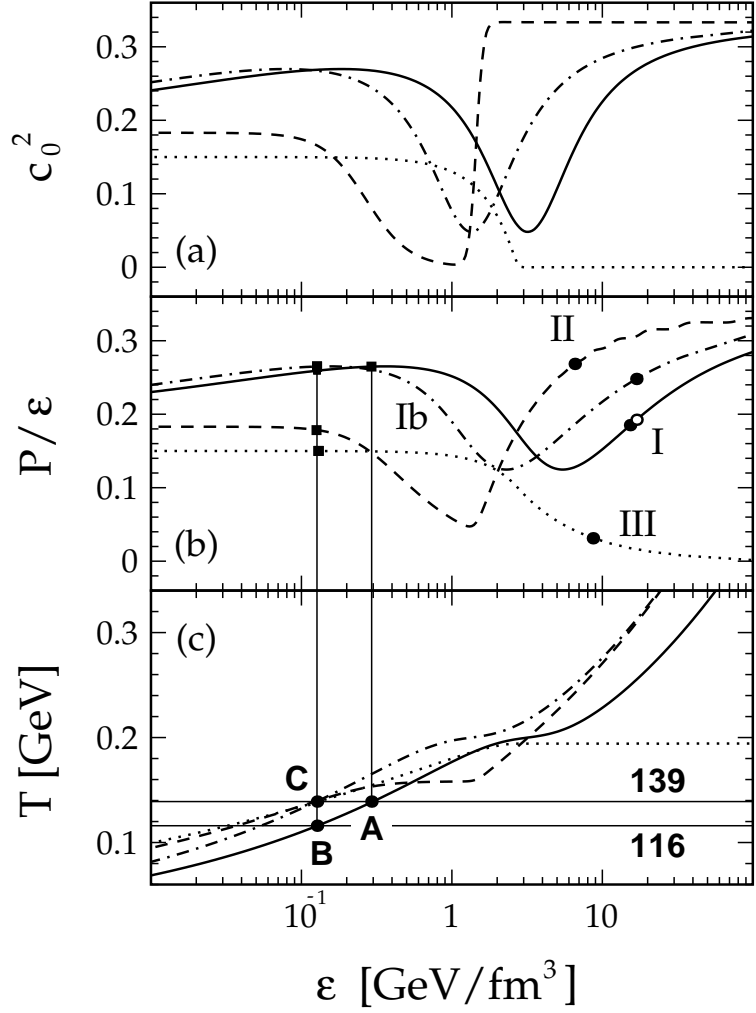,bbllx=2.0cm,bblly=3.8cm,%
bburx=20.0cm,bbury=27.7cm,width=12.0cm,clip=}
\caption{Speed of sound, $c_0^2$, ratio of pressure and energy density,
$P/\epsilon$, and temperature, $T$, as functions of $\epsilon$, 
for the equations of state EOS-I (solid lines), EOS-II (dashed lines), 
EOS-III (dotted lines), and EOS Ib (dashed-dotted lines), respectively.
The dots in plot (b) correspond for each EOS to the starting values
of $P/\epsilon$ with respect to the achieved initial maximum energy density
$\epsilon_\Delta$ at transverse position $r_\perp = 0$, (dots
correspond to $T_f = 139$ MeV whereas the open circle corresponds
to $T_f = 116$ MeV). The squares indicate the final values of $P/\epsilon$ 
at breakup energy densities, $\epsilon_f$. The dots $A$, $B$, $C$ in plot
(c) indicate the relationship between the temperature and the energy density 
at the late hadronic stage of the fireball expansion.}
\end{center}
\label{fg:fig1}
\end{figure}

After a discussion of the space-time features of the
particular fireballs we shall compare BEC of identical pion pairs
to experimental correlation functions, which have been measured by
NA44 \cite{bea98}.  

\section{The Equations of State of Nuclear Matter}

In the hydrodynamic approach, virtually any type of EOS can be 
considered when solving the relativistic
Euler-equations \cite{lan59}.  The coupled system of partial differential
equations necessary that describe the dynamics of a relativistic fluid (with
given initial conditions) contains an equation of state which we write in the
form
\begin{equation}
P(\epsilon,n) \:=\: c^2(\epsilon,n)\,\epsilon\:.  \label{eq:press}
\end{equation}
\noindent
In Eq. (\ref{eq:press}), the quantities $P$, $\epsilon$, and $n$ are the
pressure, the energy density, and the baryon density, respectively.  The
proportionality constant, $c^2$, is in general a function of $\epsilon$ and
$n$.  In the following, we shall assume that in particular the $n$ dependence
is negligible for the energy regime under consideration.  From only the
knowledge of the speed of sound, $c_o^2(\epsilon)$, one can then calculate
$c^2(\epsilon)$ by solving the integral \cite{brs98}
\vspace{-0.4cm}
\begin{equation} c^2(\epsilon) \: =\: \displaystyle{\frac{1}{\epsilon}
\int_0^\epsilon}\: c_o^2(\epsilon^\prime)\:d\epsilon^\prime\:. \label{eq:c2}
\end{equation}
\noindent
Assuming an adiabatic expansion, the temperature, $T(\epsilon)$, can be
calculated from \cite{brs98}
\begin{equation}
T(\epsilon)\:=\:T_0\:\exp\left[
\displaystyle{\int_{\epsilon_0}^\epsilon}\:
\frac{c_o^2(\epsilon^\prime)\:d\epsilon^\prime}
{(1\:+\:c^2(\epsilon^\prime))\:\epsilon^\prime}
\right] \:,
\label{eq:temp}
\end{equation}
\noindent
where $T_0=T(\epsilon_0)$ has to be specified for an arbitrary value of
$\epsilon_0$.

The first equation of state \cite{red86}, EOS-I, which we use in the 
following exhibits a phase transition to a quark-gluon plasma at a critical 
temperature $T_c$ = 200 $MeV$ with a critical energy density
$\epsilon_c$ = 3.2 $GeV/fm^3$.  The second equation of state
\cite{hun95}, EOS-II, is also a lattice QCD-based EOS which has recently 
become very popular in the field of relativistic heavy-ion physics. This
equation of state includes a phase transition to a quark-qluon plasma at
$T_c$ = 160 $MeV$ with a critical energy density $\epsilon_c$ $\approx$ 1.5
$GeV/fm^3$. The third equation of state, EOS-III, has been extracted from
the microscopic transport model RQMD \cite{sor97} under the assumption of
complete thermalization, and does {\it not} include a
transition to a QGP.  
We obtain a fourth equation of state, EOS-Ib, by changing the relationship
between $\epsilon_c$ and $T_c$ in EOS-I to $T_c (\epsilon_c = 1.35\: GeV/fm^3
)$ = 200 $MeV$.

In Fig. 1 the four equations of state are plotted in many different 
representations. The particular parametrizations for the speed of sound 
which have been used here are published in Schlei {\it et al.} 
\cite{brs98}.

\begin{figure}
\vspace*{-2.5cm}
\begin{center}\mbox{ }
\psfig{figure=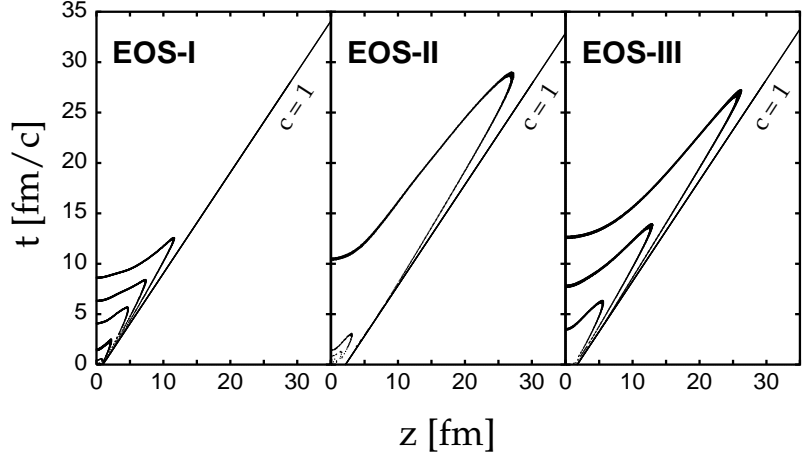,bbllx=1.0cm,bblly=14.0cm,%
bburx=21.0cm,bbury=29.7cm,width=11.0cm,clip=}
\caption{
Isothermes for the relativistic fluids governed by EOS-I, EOS-II and EOS-III
at $r_\perp = 0$, respectively.  The outer lines correspond to a temperature,
$T$ = 140 $MeV$, and each successively smaller curve represents a reduction
in temperature by $\Delta T$ = 20 $MeV$.  The lines $c=1$ represent the light
cone.}
\end{center}
\label{fg:fig2}
\end{figure}

\begin{figure}
\vspace*{-0.5cm}
\begin{center}\mbox{ }
\psfig{figure=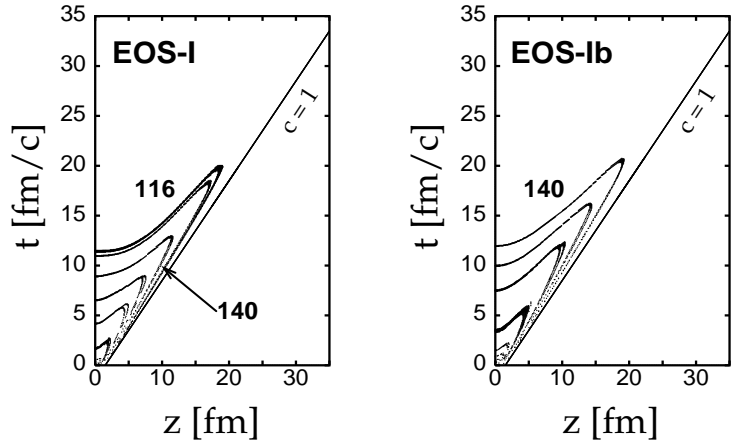,bbllx=1.0cm,bblly=2.0cm,%
bburx=21.0cm,bbury=14.0cm,width=11.0cm,clip=}
\caption{
Isothermes for the relativistic fluids governed by EOS-I, and EOS-Ib
at $r_\perp = 0$, respectively. For EOS-I the lines (beginning with the
most outer curves) correspond to temperatures, $T$ = 116 $MeV$, 120 $MeV$, 
140 $MeV$, 160 $MeV$ ... etc. For EOS-Ib, the outer lines correspond to a 
temperature,
$T$ = 140 $MeV$, and each successively smaller curve represents a reduction
in temperature by $\Delta T$ = 20 $MeV$.  
The lines $c=1$ represent the light cone.}
\end{center}
\label{fg:fig3}
\end{figure}

\section{Fits to Experimental Spectra and Freeze-Out}

In the following, we shall discuss five scenarios: we compare four
calculations using EOS-I, EOS-II, EOS-III, and EOS-Ib for the fixed freeze-out
temperature, $T_f$ = 139 $MeV$, and one calculation using EOS-I for the fixed 
freeze-out temperature, $T_f$ = 116 $MeV$.

In particular, it is assumed that due to an experimental uncertainty for the
centrality of the collision, only 90$\%$ of the total available energy and the
total baryon number have been observed. It is then possible to find initial
distributions for the four equations of state, such that one can reproduce
the single inclusive momentum spectra of 158 $AGeV$ Pb+Pb collisions. For
a freeze-out temperature, $T_f$ = 139 $MeV$, the initial conditions and a 
large number of final single inclusive momentum distributions for various 
hadron species have been published in Schlei {\it et al.} \cite{brs97_2,brs98} 
in comparison to the data measured by the NA44 \cite{nxu96,bea96} 
and NA49 \cite{jon96} Collaborations. The so far published results 
refer to calculations using EOS-I, EOS-II, and EOS-III. Results showing the 
fits of single inclusive particle momentum spectra using EOS-Ib with 
$T_f$ = 139  $MeV$ and EOS-I with $T_f$ = 116 $MeV$ will be published 
elsewhere \cite{new}.

It should be stressed, that all here discussed calculations result in single 
particle momentum distributions, which describe the corresponding data 
to the same extent very well. Although EOS-II was found in the 
calculations of hadronic transverse mass spectra to be too soft ({\it cf.} 
Schlei {\it et al.} \cite{brs97_2,brs98}), we shall use it here also for the 
calculation of Bose-Einstein correlation functions. Before we discuss BEC, 
we have to discuss briefly the thermal evolution of the various fireballs.

Figs. 2 and 3 show isothermes for the relativistic Pb+Pb fluids governed by
the different equations of state until freeze-out has been reached.
The calculation using EOS-I with $T_f$ = 139 $MeV$ leads to a fireball
of a much shorter liftime than the calculations using EOS-II and EOS-III
with $T_f$ = 139 $MeV$. This behaviour is caused by much smaller freeze-out
energy densities, $\epsilon_f$, in the calculations using EOS-II and EOS-III
compared to the calculation using EOS-I. We have $\epsilon_f$ = 0.292 [0.126
(0.130)] $GeV/fm^3$ when using EOS-I [EOS-II (EOS-III)]. A fluid that
undergoes adiabatic expansion needs more time to reach the smaller
freeze-out energy densities.

Since EOS-II and EOS-III yield similar lifetimes of the fireball, in
the following we attempt to increase the lifetime of the system which
is governed by EOS-I. This can be acchieved by (a) using a smaller
freeze-out temperature, $T_f$ = 116 $MeV$, or (b) by hardening the EOS,
i.e., using EOS-Ib instead of EOS-I (without changing $T_f$ = 139 $MeV$). 
For the latter two cases, we obtain $\epsilon_f$ = 0.127 $GeV/fm^3$ ({\it cf.} 
Fig. 1 (c) and Fig. 3).

\begin{figure}
\vspace*{-1.8cm}
\begin{center}\mbox{ }
\psfig{figure=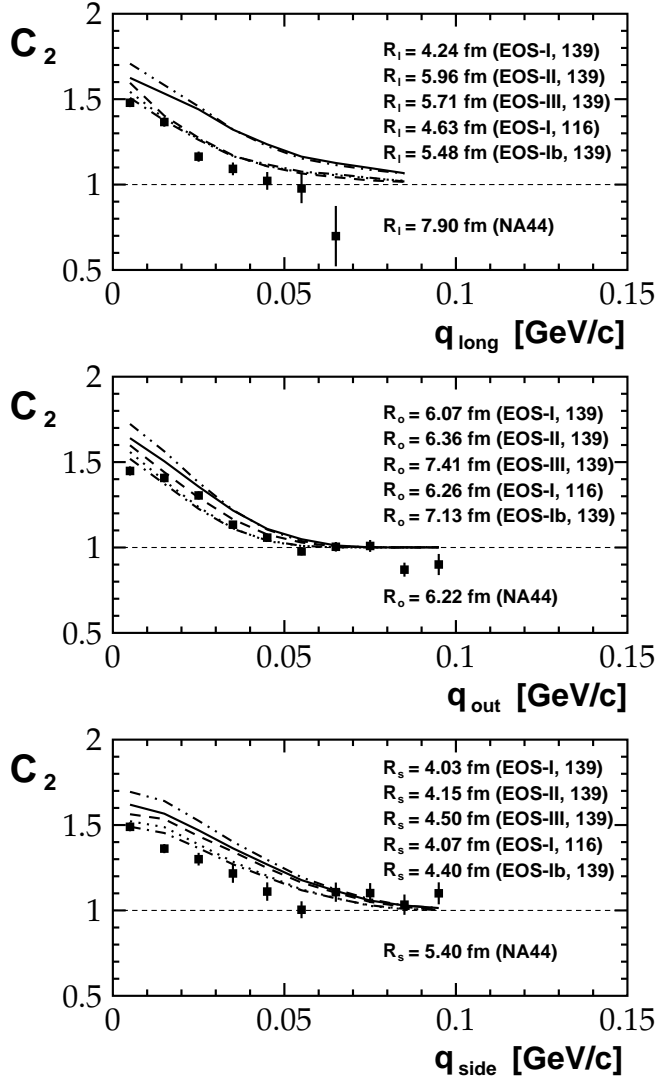,bbllx=1.0cm,bblly=1.0cm,%
bburx=21.0cm,bbury=29.7cm,width=11.4cm,clip=}
\caption{Projections of BEC functions for $\pi^+\pi^+$ pairs emerging from 
the low $p_\perp$ (horizontal and vertical) acceptance setting of the NA44 
detector $^8$. The data points are data taken by the NA44 Collaboration $^8$. 
The solid [dashed [dotted [dashed-dotted]]] lines correspond 
to the calculations using EOS-I [EOS-II [EOS-III [EOS-Ib]]] with $T_f$ = 139
$MeV$, and the double dotted-dashed lines correspond to the calculation
using EOS-I with $T_f$ = 116 $MeV$. The values of the inverse width 
parameters, $R_i$ ($i=l,o,s$), have been obtained from the Gaussian
Bertsch-Pratt parametrization (see text).}
\end{center}
\label{fg:fig4}
\end{figure}

\section{Bose-Einstein Correlations}

The lifetime of the fireball is, e.g., reflected in Bose-Einstein correlations
of identical pion pairs \cite{brs97}. Within the Bertsch-Pratt variables 
\cite{ber88}, a very sensitive quantity -- regardless of whether one is 
interested in transverse expansion \cite{brs92} and BEC, or in the role of 
resonance decay \cite{brs93} in BEC -- has always been the longitudinal 
projection of the two-pion correlation function, $C_2(q_{long})$. 
Fig. 4 shows projections of BEC functions for $\pi^+\pi^+$ pairs emerging from 
the low $p_\perp$ (horizontal and vertical) acceptance setting of the NA44 
detector \cite{bea98}. The data points are data taken by the NA44 
Collaboration \cite{bea98}.

It must be stressed, that in the calculations it was necessary to fully
simulate the experimental detector acceptance. I have checked, that if one 
{\it does not follow} the experimental prescription in generating the 
correlation functions one could obtain errors for the inverse widths and the
overall strengths of the correlation functions in the order of 25\%.

Consistent with the expectation, the calculations using EOS-II, EOS-III,
and EOS-Ib give similar lifetimes and therefore sufficiently large
longitudinally expanded fireballs (see Figs. 2,3). In particular, these 
calculations give -- in addition to the excellent description 
\cite{brs97_2,brs98,new} of hadronic 
single inclusive momentum spectra (except for EOS-II) -- a very nice 
reproduction of the pionic NA44 BEC data. It should be stressed here, that a 
freeze-out temperature $T_f$ = 139 $MeV$ was {\it sufficient} to achieve this 
agreement.

Also consistent with expectation is that the calculation using EOS-I with 
$T_f$ = 139 $MeV$ gave a too small longitudinally expanded fireball 
(see above, and Fig. 2). On the contrary, it is on first sight surprising 
that a reduction of the freeze-out temperature to $T_f$ = 116 $MeV$ does not 
lead to large enough longitudinal extension of the fireball in the calculation 
which uses EOS-I. The reason for this result is the following: 
a freeze-out temperature reduction leads to a larger lifetime of 
the {\it direct} fireball, but the relative fraction of present resonance 
decay contributions is reduced (by about 30\%); i.e., the resonance halo is 
reduced in size. Hence, the {\it total} fireball, which is a superposition
of the direct (or thermal) fireball and the resonance halo remains
moreless unchanged in size.

In conclusion, by inspecting Fig. 1(c) we can see that only those equations
of state which go through point $C$ in the figure reproduce the experimental 
data on Bose-Einstein correlations good enough. Changing the freeze-out
temperature shows hardly any effect. Therefore, the measurements
of Bose-Einstein correlations tell us about which relationship between
temperature and energy density is neccessary for a valid choice of an
equation of state in the calculations. 
Unfortunately, from the above made considerations it must be noted, 
that two-particle BEC {\it seen by itself} cannot be used as a tool 
to determine a possible phase-transition to a QGP, because the BEC show no 
sensitivity to the structure of the EOS.

\section*{Acknowledgments}

I would like to thank the organizers for their invitation to
present these results at the ``Correlations and Fluctuations '98''
workshop in M\'atrah\'aza, Hungary. In particular, I am grateful
for discussions with Drs. D. Strottman and M. Gyulassy. Special thanks
goes to Drs. J.P. Sullivan and H.W. van Hecke for explaining and
providing the NA44 Bose-Einstein data to me. 
This work has been supported by the U.S. Department of Energy.

\end{document}